\documentclass[11pt]{article}
\usepackage{graphicx} 
\usepackage{amsfonts, amsbsy, amssymb}

\begin{document}


\newtheorem{theorem}{Theorem}
\newtheorem{proposition}{Proposition}
\newtheorem{corollary}{Corollary}
\newtheorem{lemma}{Lemma}

\def \R{{\mathbb R}}
\def \T{{\mathbb T}}
\def \Z{{\mathbb Z}}
\def \eps{\varepsilon}
\def \u {\boldsymbol u}

\title{Friedman vs Abel equations:
A connection unraveled}
\author{A.V. Yurov, V.A. Yurov }

\maketitle

\maketitle
\begin{abstract}
We present  connection between Einstein-Friedmann equations for
the models of universe filled with scalar field and the special
form of Abel equation of the first kind. This connection works in
both ways: first, we show how, knowing the general solution of the
Abel equation (corresponding to the given scalar field potential)
one can obtain the general solution of the Friedman Equation (and
use the former for studying such problems as existence of
inflation with exit for particular models). On the other hand, one
can invert the procedure and construct the B\"{a}cklund
auto-transformations for the Abel equation.

\end{abstract}

\subsection*{1. Introduction}

Starting out from the classical work of Abel \cite{Abell}, the Abel equation has seen a lot of extensive studies and applications \cite{Appel}, \cite{Liu} (see
also \cite{111}, \cite{222}). The reason for later lies in the fact that Abel equations rather frequently appear in the process of
reduction of order for many second (and higher) order families
\cite{Kamke}, \cite{ops}, and hence are often found in the
modeling of real problems in varied areas.

In this work we present but another new field of application of the Abel
equation: as we shall see, the general solution of the cosmological
Einstein-Friedmann equations for the universe filled with scalar
field for a given potential can be expressed via the general solution of
the Abel equation of the first kind.

Let us start by writing the Friedmann equations describing
homogeneous isotropic universe (in
Friedmann-Lemaitre-Robertson-Walker metric):

\begin{equation}
\left\{\begin{array}{ll} \displaystyle{\ddot \phi+3H\dot \phi+\frac{dV}{d\phi}=0}
\\
\displaystyle{H^2=\frac{1}{2}\dot\phi^2+V-\frac{k}{a^2}}
\end{array}\right.
\label{fried}
\\
\end{equation}
where $a$ is a scale factor, $\phi$ - scalar field, $V$ is the
scalar field's self-action potential, $H$ - Hubble constant $H=\dot
a/a$, $k=0, \pm1$ and we have incorporated the rescaling $\frac{8\pi
G}{3}=c=1$.

The general study of the dynamics of the universe for a given
self-action potential is an exceptionally difficult mathematical
problem. Nevertheless, this paper is dedicated to the task of
finding out a regular procedure of construction of the system
(\ref{fried})'s general solution for a given potential. By
definition, the general solution has to have two arbitrary
integration constants: $\phi=\phi(t;t_0,C)$ (constant $t_0$ is
connected to the translation invariance of (\ref{fried}))
Restricting ourselves to the case of flat space-time $k=0$, it
would be enough to define the general solution for $\phi$, since
for such model it is possible to exclude the Hubble constant from
the first equation of (\ref{fried}) and rewrite it as
$$\ddot \phi\pm
3\sqrt{\frac{1}{2}\dot\phi^2+V}\dot \phi+\frac{dV}{d\phi}=0
$$
Scale factor $a(t;t_0,C)$ then would be defined by simple integration of
the second equation of (\ref{fried}), the process that will give
birth to the second (multiplicative) integration constant,
allowing for the scale factor rescaling.

\noindent\textbf{Remark 1.}  For any twice differentiable scale factor function $a=a(t)$ and for
any $k$ it is possible to find out the corresponding scalar field
that will provide the necessary dynamics. This can be done via the
following formulas:
\begin{equation}
\begin{array}{cc}
\displaystyle{\phi(t)=\phi_0\pm \sqrt{\frac{2}{3}}\int
dt~\frac{\sqrt{\dot a^2-a\ddot a+k}}{a}}
\\
\\
\displaystyle{V(t)=\frac{a\ddot a+2\dot a^2+2k}{3a^2}.}
\\
\\
\end{array}
\label{param}
\end{equation}
Equations (\ref{param}) describe the function $V=V(\phi)$ in the
parametric form defined for any given function $a(t)$. Of course,
this procedure has nothing to do with the searches for the general
solution of (\ref{fried}). First of all, the described procedure
might at best help to introduce the first of the constants
($\phi_0$, obtained via the integration of the first equation of
(\ref{param})), but not the second one. It is by all means
possible to choose the scale factor depending on the arbitrary
amount of undefined constants $\{C_1,...,C_N\}$ so that they will
appear in the solution $\phi=\phi(t,C_1,...,C_N)$. But then, as
follows from (\ref{param}), the potential $V$ will depend on those
constants as well.

It's second (and worst from the physical point of view) fault lies
in the fact that the potential has to be derived from the known
scale factor, and not the other way around - in this sense, it can
be called the ''inverse'' problem. As easy as the ''inverse''
problem goes, the ''direct'' problem is way more difficult - it
would be enough to say that in most cases it is being solved only
by the procedures of the numeric calculations. However, as we will
show below, it is possible to reduce it to one of the most
well-developed problem: to finding the solution of the particular
Abel equation of the first kind.

\subsection*{2. The functional of full energy}

\medskip\noindent
In the rest of the article we'll restrict ourselves to the case
$k=0$ as the one best fit to describe the observed universe.
Following the ideas of the \cite{chervon} (see also \cite{YAY}), let us introduce the
functional of full energy (hamiltonian) $W$:

\begin{equation}
W=\frac{1}{2}\dot\phi^2+V(\phi) \label{ham}.
\end{equation}
It is easy to see that, using (\ref{ham}) and assuming $k=0$
system (\ref{fried}) can be rewritten as:

\begin{equation}
\begin{array} {cc}
\displaystyle{\frac{dW}{d\phi}=-3H\dot\phi}
\\
H=\pm\sqrt{W}
\end{array}
\label{newfr}
\\
\\
\end{equation}
Moreover, knowledge of the function $W=W(\phi)$ provides an
elementary way of finding all other quantities ($\phi=\phi(t)$,
$a=a(t)$, $V=V(\phi)$). If $W\neq0$ (cf. Remark 2), $\phi(t)$ can be derived from the ordinary
differential equation
\begin{equation}
\frac{d\phi}{dt}=\mp\frac{1}{3}\frac{W'(\phi)}{\sqrt{W(\phi)}}
\label{newphi}
\\
\\
\end{equation}
and potential $V$:
\begin{equation}
V(\phi)=W(\phi)-\frac{1}{18}\frac{(W'(\phi))^2}{W(\phi)}.
\label{newV}
\\
\\
\end{equation}
As for the scale factor $a(t)$ = exp $(\int H(t)dt)$, it can be
obtained by substituting $\phi(t)$ from (\ref{newphi}) into
(\ref{newfr}) and consequent integration.
\newline\newline
\noindent\textbf{Example 1.} Choose $W(\phi)=\lambda \phi^4/4$,
$\lambda>0$. According to (\ref{newV}), this hamiltonian
corresponds to the potential with the spontaneously broken
symmetry (note, that the value of constant $\mu$ in (\ref{exV})
isn't arbitrary: $\mu^2=4\lambda/9$)
\begin{equation}
V(\phi)=\frac{\lambda\phi^4}{4}-\frac{\mu^2}{2}\phi^2, \label{exV}
\end{equation}
whereas the dynamical variables  would be:

\begin{equation}
\begin{array}{cc}
\displaystyle{a(t)=a_0~ \rm
exp\left(\frac{3\phi_0^2}{8}\left[1-e^{\pm\frac{4\sqrt{\lambda}(t-t_0)}{3}}\right]\right)}
\\
\displaystyle{\phi(t)=\phi_0~\rm
exp\left(\pm\frac{2\sqrt{\lambda}}{3}(t-t_0)\right)}
\\
\displaystyle{H(t)=\mp\frac{\sqrt{\lambda}~\phi_0^2}{2}~\rm
exp\left(\pm\frac{4\sqrt{\lambda}(t-t_0)}{3}\right)}.
\end{array}
\label{dyna}
\\
\\
\end{equation}
Here the variables with zero subscript correspond to current
($t=t_0$) values of scalar field and scale factor.

Integration of (\ref{newphi}) gives rise to the constant $t_0$,
which appears because of the translational invariance of
(\ref{fried}) and can be assimilated by the translation $t\to
t+{\rm const}$. In order for the solution $\phi=\phi(t;t_0,C)$ to
be general it has to contain the second independent constant $C$.
This constant can be obtained through the following calculations:
consider (\ref{newV}) as a differential equation w.r.t. variable
$W(\phi)$ with the given $V(\phi)$. General solution of this
equation contains the thought after integration constant $C$.
Thus, we can now formulate the following

\begin{proposition}
If for a given $V(\phi)$ the general solution of equation
(\ref{newV}) is $W=W(\phi,C)$, the general solution of the
Friedmann equation (\ref{fried}) $\phi(t;t_0,C)$ exactly
corresponds to the general solution of (\ref{newphi}).
\end{proposition}
Hence, the problem is reduced to the task of finding the general
solution of  (\ref{newV}) for a given  $V(\phi)$.

\noindent\textbf{Remark 2.} The case $W=0$ shall be considered
separately. It is easy to verify that for any given $V\leq0$ one gets $H=0$
($a(t)=a_0={\rm const}$ - stationary universe) and the general
solution of the (\ref{fried}) contains the single integration
constant:
\begin{equation}
\left\{\begin{array}{ll} \displaystyle{\int\frac{d\phi}{\sqrt{-2V(\phi)}}=\pm\left(t-t_0\right)} ;& V\neq0 \\\
\\
\displaystyle{\phi=\phi_0} ;& V=0
\end{array}
\right.
\end{equation}
So this model is meaningful for the non-positive potential $V(\phi)$
only \cite{negative}. It is also interesting to note that the parameter of equation of state $w=p/\rho=\infty$
where $\rho$ is the density and $p$ is the pressure of scalar
field.

In what follows we'll restrict ourselves to the case $W\neq0$.

\subsection*{3. Main Theorem}

The main result of this paper lies in the following theorem:

\begin{theorem}
Let $x=3\sqrt{2}\phi$, $\chi=\rm{ln} |V|$, $\kappa=\pm1$. For a
given $V(\phi)$ the corresponding hamiltonian $W=W(x,C)$ is defined as (cf. Remark 2):
\begin{equation}
W(x,C)=V(x)\left(\frac{\left(y+\sqrt{y^2-1}\right)^2+1}{1-\left(y+\sqrt{y^2-1}\right)^2}\right)^2
\label{theW}
\end{equation}
where $y=y(x,C)\neq\pm1$ is a general solution of Abel equation of
1st kind:
\begin{equation}
y'=-\frac{1}{2}\left(y^2-1\right)\left(\kappa-\chi'y\right).
\label{abel}
\end{equation}
Moreover, the special case $V=0$ occurs if and only if $y=\pm1$ and the hamiltonian W has the form:
\begin{equation}
W=C e^{\kappa x}
\end{equation}
\end{theorem}

The proof of the theorem can be performed by the direct
calculations.
\newline

\noindent\textbf{Remark 3.} (\ref{theW}) defines a family of
solutions of (\ref{fried}), parameterized by the constant $C$.
Substituting $W(x,C)$ in (\ref{newphi}) after the integration  one
will obtain $\phi=\phi(t;C,t_0)$, where $t_0$ is the second
integration constant, connected to the invariance of the scalar
field relative to translations $t\to t+\rm const$. In other words,
the suggested algorithm indeed allows to find the general solution
of (\ref{fried}), which is a main result of the paper.
\newline
\newline
\noindent\textbf{Remark 4.} For the special case $y=\pm1$ (i.e. $V=0$) the general solution of the
(\ref{fried}) has the form
$$
\phi(t)=\phi_0 \pm\frac{\sqrt{2}}{3}\log\left(t-t_0\right),
$$
the scale factor is
$$
a(t)=a_0\left(t-t_0\right)^{1/3}
$$
and the case corresponds to the ''stiff'' equation of state with $w=1$.
\newline
\newline
\noindent\textbf{Remark 5.} If $\chi'\neq \rm const$ (cf. Ch. 4, example D) the eq. (\ref{abel}) has two fixed
points $y=\pm 1$. It means that for any initial $y_0=y(x_0,C)\ne
\pm1$ the solution $y(x,C)\ne\pm1$ for all values of $x$ excluding
probably $x=\pm\infty$.

 The general form of Abel equation of first kind
is:
\begin{equation}
y'=\sum_{\eta=0}^3~f_\eta~ y^\eta, \label{abelG}
\end{equation}
where, in our case:
\begin{equation}
f_0=\frac{\kappa}{2},~f_1=-\frac{1}{2}\chi',~f_2=-\frac{\kappa}{2}=-f_0,~f_3=\frac{1}{2}\chi'=-f_1.
\label{smile}
\end{equation}

As known (cf., for example \cite{222}, \cite{Kamke}), if $f_1$ is continuous, $f_2$ and $f_3$ are continuously
differentiable  and $f_3\ne 0$ then one can represent the
equation (\ref{abel}) in normal form:
\begin{equation}
\eta'=\eta^3+J(x),
\label{normal}
\end{equation}
where
\begin{equation}
\begin{array}{cc}
\displaystyle{ y=\omega(x)\eta(\xi)+\frac{\kappa
V(x)}{3V'(x)}},\qquad
\omega(x)=\frac{1}{\sqrt{V(x)}}\exp\left(-\frac{1}{6}\int^x\frac{V(z)}{V'(z)}dz\right),\\
\\
\displaystyle{
\xi=\frac{1}{2}\int\frac{V'(x)\omega^2(x)}{V(x)}dx,\qquad
J=\frac{2\kappa\left(9V''-V\right)V^2}{\left(3V'\omega\right)^3}.}
\end{array}
\label{substi}
\end{equation}

\noindent\textbf{Example 2.} Let consider  the popular
cosmological model with quadratic potential
\begin{equation}
 V(\phi)=\frac{m^2\phi^2}{2},
 \label{quadro}
\end{equation}
In a quantum field theory this model describes noninteracting
massive scalar particles. We shall carefully consider  this model
in the Sec.5. Now one can use the substitution (\ref{substi}) for
the $\phi>0$ ($x>0$). In this case
\begin{equation}
\begin{array}{cc}
\displaystyle{ \omega(x)=\frac{6}{m x}{\rm e}^{-x^2/24},\qquad
J(x)=\frac{\kappa m^3}{23328}x^4\left(x^2-18\right){\rm
e}^{x^2/8},}\\
\\
\displaystyle{ \xi=\frac{3}{2m^2}\left[{\bf {\rm
Ei}}\left(1,\frac{x^2}{12}\right)-\frac{12}{x^2}{\rm
e}^{-x^2/12}\right].}
\end{array}
\label{subs-1}
\end{equation}
\noindent\textbf{Remark 6.} In the case of polynomial potentials
$$
V=\frac{\lambda\phi^n}{n}=\frac{\lambda x^n}{18^{n/2}n},
$$
with positive coupling $\lambda>0$, the Abel  equation of 1st kind
can be transformed into the particular case of the Abel  equation
of 2st kind. In this case one gets $\chi'=n/x$. Assuming $\left|y_0\right|>1$ and $\rm sign(y_0)=-\kappa$, lets consider $y$ as
the independent variables and $x=x(y)$ as a solution under the
question. Introducing new function $P=P(y)=\kappa x(y)-ny$ and
substituting it into the (\ref{abel}) one get
\begin{equation}
PP'=F_1(y)P+F_0(y), \label{abel2kind}
\end{equation}
with
$$
F_1(y)=-n-\frac{2}{y^2-1},\qquad F_0(y)=-\frac{2ny}{y^2-1}.
$$
Lets define
$$
P=u(y)+F(y),\qquad F(y)=\int^y_0 F_1(z)dz=-ny+\log\left|\frac{y+1}{y-1}\right|.
$$
Then the equation (\ref{abel2kind}) will be reduced to
\begin{equation}
(u+F)u'=F_0. \label{abel2kind-1}
\end{equation}
Finally, if we'll assume that $\left|y_0\right|\ge 1$ then $F_0\ne 0$ and one can
introduce a new independent variable
$$
\xi=\int^y_0 F_0(z) dz=-n\log\left(y^2-1\right), \qquad u(y)=\eta(\xi).
$$
with whom the equation (\ref{abel2kind-1}) will has the normal form
$$
\left(\eta+F\right)\eta'=1.
$$

Another form of the Abel equation of 1st kind (\ref{abel}) can be
obtained if we know at least one of it's exact solutions. For
(\ref{abel}) there are two such solutions $y=\pm 1\equiv K$.
Calculating the function
$$
E(x)=\exp\int\left(3f_3 K^2+2f_2K+f_1\right)dx=V(x){\rm
e}^{-\kappa Kx},
$$
and using the substitution
$$
y=K+\frac{E(x)}{z(x)},
$$
one get
\begin{equation}
z'+\frac{\Phi_1}{z}+\Phi_2=0, \label{abel-2}
\end{equation}
where
$$
\Phi_1=\frac{1}{4}\left(V^2\right)'{\rm e}^{-2\kappa Kx},\qquad
\Phi_2=\frac{1}{2}{\rm e}^{-\kappa Kx}\left(3KV'-\kappa V\right).
$$
For the case
$$
V=V_0{\rm e}^{\kappa K x/3}
$$
$\Phi_2=0$ and the equation (\ref{abel-2}) can is exactly solvable. This example will be thoroughly examined in the next section.
\newline
\newline
\noindent\textbf{Remark 7.} There have been many works recently that treat either modified
gravity or $f(R)$-gravity \cite{modgrav-1},  \cite{modgrav-2},
 \cite{modgrav-3}, \cite{modgrav-4}, \cite{modgrav-5}. All these models
suggest an alternatives for the origin of dark energy. It may be naturally
expected that gravitational action contains some extra terms which
became relevant recently due to the significant decrease of the
universe curvature.

The $f(R)$ models can in general be describes via the action
$$
S=\frac{1}{k^2}\int d^4x\sqrt{-g}\left[R+f(R)\right],
$$
where $f(R)$ is the proposed additional term. After the introduction of the
Friedmann-Lemaitre-Robertson-Walker metric
\begin{equation}
ds^2=-dt^2+a^2(t)\sum_{i=1}^3 \left(dx^i\right)^2, \label{FRW}
\end{equation}
one gets new equations, naturally differing from the generic
Friedmann equations (\ref{fried}). Clearly, the general solutions problem in this new framework gets even more difficult then before.

However, one can still reduce any given $f(R)$ model to a special usual
Einstein-Friedmann universe filled with scalar field with non
minimal coupling. In fact, letting $a(t)$ be the particular solution
of some $f(R)$-model in metric (\ref{FRW}) and substituting it into the (\ref{param}) one will obtain the function $V=V(\phi)$ in
the parametric form defined for the given scalar factor.
Therefore, we end up with the following
\begin{proposition}
For any solution $a(t)$ of any  $f(R)$-model in metric
(\ref{param}) there exist an Einstein-Friedmann model with scalar field
having exactly the same scale factor.
\end{proposition}

In other words, one can reduce $f(R)$-models to usual Friedmann equation
(\ref{fried}) in order to find their general solutions.

\subsection*{4. Some Examples}

 Let's consider a couple
of examples of potentials that are frequently used in cosmological
studies.
\newline

a)
$$
V(\phi)=\frac{\lambda \phi^n}{n}.
$$
Here we have the following set of functions $f_\eta$:
$$
f_3=-f_1=\frac{n}{2x},~f_2=-f_0=-\frac{\kappa}{2}.
$$

b)
$$
V(\phi)=\frac{\lambda \phi^4}{4}+\frac{m^2}{2}\phi^2.
$$
This potential describes the popular field model of scalar
particles with the coupling constant $\lambda>0$ and the mass $m$.
$$
f_3=-f_1=\frac{2(\lambda x^2+18m^2)}{x(\lambda x^2+36m^2)}.
$$

c)
$$
V=\Lambda=\rm const.
$$
$$
f_3=-f_1=0.
$$
if $\phi=\rm const$ this case corresponds to the model with the
cosmological constant.
\newline

d)
$$
V=V_0 e^{6\sqrt{2}\alpha\phi},~ \alpha=\rm const.
$$
$$
f_3=-f_1=\alpha.
$$

Let us now consider the cases c) and d) in details.

If $V=\Lambda=\rm const$, equation (\ref{abel}) has the solution:
\begin{equation}
y(x,x_0)=\frac{e^{\kappa(x-x_0)}+1}{e^{\kappa(x-x_0)}-1},
\end{equation}
and the functional $W$ has the form:
\begin{equation}
W(x,x_0)=\Lambda~ \cosh^2 \left(\frac{\kappa}{2}(x-x_0)\right).
\end{equation}
After substitution into (\ref{newphi}) we obtain the following
differential equation:
\begin{equation}
\dot
\phi=\pm\kappa\sqrt{2\Lambda}~\sinh\left(\frac{3\kappa}{\sqrt{2}}(\phi-\phi_0)\right).
\label{phidot}
\end{equation}
 The general solution of
(\ref{phidot}) $\phi(x;x_0,t_0)$ will be parameterized by two
arbitrary constants $x_0$, $t_0$ and will have a form:
\begin{equation}
\phi=\phi_0+\frac{\sqrt{2}}{3\kappa}~ {\rm
arccosh}\left(\cot~\left[3\sqrt{\Lambda}\left|t-t_0\right|\right]\right) \label{phi}
\end{equation}
The scale factor
\begin{equation}
a(t)=a_0\left[\sinh\left(3\sqrt{\Lambda}\left|t-t_0\right|\right)\right]^{\pm
1/3}. \label{at}
\end{equation}

Note, that if $t\to\infty$ then $\phi\to\phi_0$ (see (\ref{phi}))
so the density $\rho\to\Lambda$ and the pressure $p\to-\Lambda$.
In the case of positive power in (\ref{at}) we obtain the
well-known de-Sitter (dS) solution as $t\to\infty$:
$$
a(t)\to \frac{a_0}{2^{1/3}}{\rm e}^{\sqrt{\Lambda}t},
$$
while the negative power case contribute the solution with the so called Big Rip
singularity at $t=t_0$ which means that one deals with the phantom
cosmology here \cite{BR-1}, \cite{BR-2}, \cite{BR-3}, \cite{BR-4}.

For the case d)  equation (\ref{abel}) takes the form
\begin{equation}
y'=\alpha(y-1)(y+1)(y-s), \label{ab-22}
\end{equation}
where $s=\kappa/(2\alpha)$. This equation has  three  fixed
points: $y=\pm 1$ and $y=s$. For simplicity one choose $s>1$.
Therefore if the initial value $1<y_0<s$ then this will be the
case for $y(x,C)$ for any $x$. The solution of the (\ref{ab-22})
has the form
\begin{equation}
\left(\frac{y+1}{y-1}\right)^s\frac{(y-s)^2}{y^2-1}=C{\rm
e}^{2\alpha(s^2-1)x}, \label{soll}
\end{equation}
where $C>0$ is an integration constant. One can see that
$$
\begin{array}{l}
y\to s\qquad\qquad {\rm at}\qquad\qquad x\to+\infty\\
y\to 1\qquad\qquad {\rm at}\qquad\qquad x\to-\infty
\end{array}
$$
for $\alpha<0$ and
$$
\begin{array}{l}
y\to s\qquad\qquad {\rm at}\qquad\qquad x\to-\infty\\
y\to 1\qquad\qquad {\rm at}\qquad\qquad x\to+\infty
\end{array}
$$
for $\alpha>0$.

\subsection*{5. The $\frac{m^2\phi^2}{2}$ model: inflation and slow-rolling approximation}

The Abel representation (\ref{abel}) of Friedmann equation (\ref{fried}) can be extremely useful even in the cases when one cannot find it's exact solution. To demonstrate this let us consider the popular cosmological model with the quadratic potential (\ref{quadro}). It is known that Friedmann equation with this potential is non-integrable. Same goes for the Abel equation. However, by studying this model it is still possible to draw some interesting conclusions, namely: that the equation (\ref{fried}) may result in inflation with natural exit from it.

The standard approach to this task (cf., for example, \cite{textb}) would be to use the slow-rolling approximation, i.e. assume that
\begin{equation}
K=\frac{{\dot\phi}^2}{2}\ll |V|. \label{slow-roll}
\end{equation}
If this is true, the equation of state is $p\sim -\rho$ and one gets
inflation. For polynomial potential (akin to the one in question)
condition (\ref{slow-roll}) may be valid if $\phi\gg 1$. It is
then assumed that during the inflation process the kinetic term
$K$ increases until the slow-rolling approximation
(\ref{slow-roll}) is no longer applicable, which heralds the (spontaneous) natural exit from the inflation. This reasoning, being rather simple as an idea proved to be quite a challenge when it came to the rigorous proofs in particular cases. In the last 14 years this difficult problem has drawn a lot of interest and attention. The investigations conducted in these field (see \cite{int-1}, \cite{int-2},
\cite{int-3}, \cite{int-4} and list of references) have led to the following
conclusions:
\newline
1. The idea of slow rolling is true. Inflation does indeed occur
under an extremely broad range of self-acting potential, and there
is hence no need to fix a certain form of the potential  to obtain
an inflationary universe.
\newline
2. The exit from inflation, on the contrary, turned out to be a problem.
For many model potentials the universe never stops inflating. In general, the
exit has to be achieved only by the means of fine-tuning or, in other words, by
the parameter fitting.

In this Section, by using the Abel equation we show that the popular
non-integrable model with quadratic potential (\ref{quadro}) is
free from the later problem, which lies in total accordance with the conclusions achieved so far.

Let us start by introducing the quantity $\theta^2(y)$ (cf. (\ref{theW})):
\begin{equation}
\theta^2(y)=\left(\frac{\left(y+\sqrt{y^2-1}\right)^2+1}{1-\left(y+\sqrt{y^2-1}\right)^2}\right)^2.
\label{thetta} \end{equation}
For $y\ge 1$ this is the monotonously decreasing function of $y$, such that:
$$
\lim_{_{y\to 1}}\theta^2(y)=+\infty,\qquad \lim_{_{y\to
\infty}}\theta^2(y)=1.
$$
The plot of $\theta^2(y)$ is represented on the Fig.1.
\begin{figure} \begin{center}
  \includegraphics[angle=270, scale=0.35]{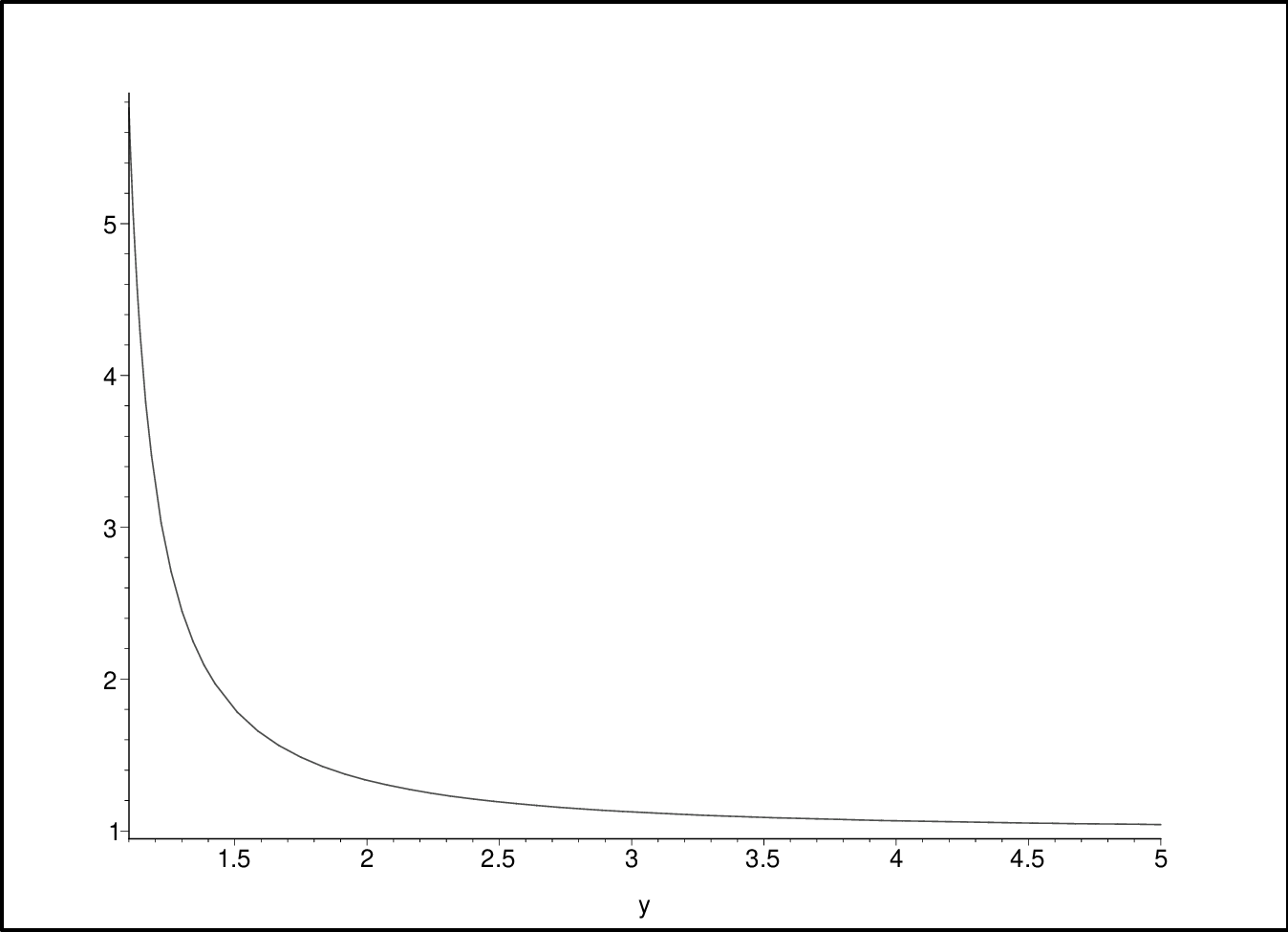}
  \caption{The plot of $\theta^2(y)$.}
\end{center}
\end{figure}
The slow-rolling approximation (\ref{slow-roll}) can be
rewritten in the form
\begin{equation}
\frac{K}{|V|}=\theta^2(y)-1\ll 1. \label{slr}
\end{equation}
For the $y\to\infty$ we have $\theta^2(y)\sim 1+1/y^2$ therefore the
approximation (\ref{slr}) will be valid if $y\gg 1$.

The inflation take place whenever ${\ddot a(t)}/{a(t)}>0$ (which is identical to condition $\rho+3p<0$). This will be the case if $\theta^2(y)<3/2$ or
$y>\sqrt{3}$. The pressure will be negative if $\theta^2(y)<2$ or
$y>\sqrt{2}$. All these results are presented in Tabl. 1.

\begin{table}
    \begin{tabular}{|c|c|c|c|c}\hline ${}$ & Slow-rolling  ($\frac{\dot\phi^2}{2}\ll |V|$)
    &Inflation ($\rho+3p<0$) & Negative pressure   \\\hline
 I:\,\, $1\ll y_*<y<\infty$& yes& yes &yes
 \\\hline II:\,\,$\sqrt{3}<y<y_*$
&no&yes&yes
\\\hline
III:\,\,$\sqrt{2}<y<\sqrt{3}$ &no&no&yes
\\\hline
IV:\,\,$y<\sqrt{2}$ &no&no&no
\\\hline
\end{tabular}
\caption{}
\end{table}



Now consider the Abel equation (\ref{abel}) with the quadratic
potential (\ref{quadro}), i.e. with $\chi=2/x$. For simplicity let us
choose $\kappa=+1$. The equation (\ref{abel}) takes the form:
\begin{equation}
y'=-\frac{1}{2}\left(y^2-1\right)\left(1-\frac{2y}{x}\right).
\label{Abb}
\end{equation}
\noindent\textbf{Remark 8.} Since  $y=\pm 1$ are fixed points of
the equation (\ref{Abb}) the range of an arbitrary solution of
(\ref{Abb}) $y(x,C)$ (in this chapter for the sake of simplicity
we'll refer to the solutions of (\ref{Abb}) as just $y(x)$,
omitting the constant $C$) will be splitted  into the three
distinct regions: $I_1=\{y~|y>+1\}$; $I_2=\{y~|-1<y<+1\}$; and
$I_3=\{y|y<-1\}$. Curve $y(x)$ can cross the lines $y=\pm 1$ in
two possible cases: (i) at $x=0$, which is a singular point of
(\ref{Abb}) and (ii) if $y(x)$ is singular in at least one point
$x=x_s$.  In the framework of the discussed applicability of the
Abel equations to the cosmological models with real scalar field,
the most interesting regions would be $I_1$ and $I_3$. In this
work we'll concentrate on studying $I_1$.

Before we actually start tackling the equation (\ref{Abb}), it would be worthwhile to spend some time on analyzing the easier but still somewhat similar Riccatti equation :
\begin{equation}
y'=-\frac{1}{2}\left(y-1\right)\left(1-\frac{2y}{x}\right).
\label{Ric}
\end{equation}
Similarly to (\ref{Abb}), equation (\ref{Ric}) also has a fixed point at $y=1$ as well as the singular point at $x=0$. However, contrary to (\ref{Abb}), this equation can be solved explicitly; it's general solution has a form:
\begin{equation}
y(x,C)=\frac{x+2+C{\rm e}^{x/2}}{2+C{\rm e}^{x/2}}, \label{R1}
\end{equation}
where $C$ is an integration constant.
(\ref{R1}) is obviously regular whenever $C>0$, and $C<0$ implies existence of singularity at a certain point $x_s$ (see Fig. 2).
\begin{figure} \begin{center}
  \includegraphics[angle=270, scale=0.25]{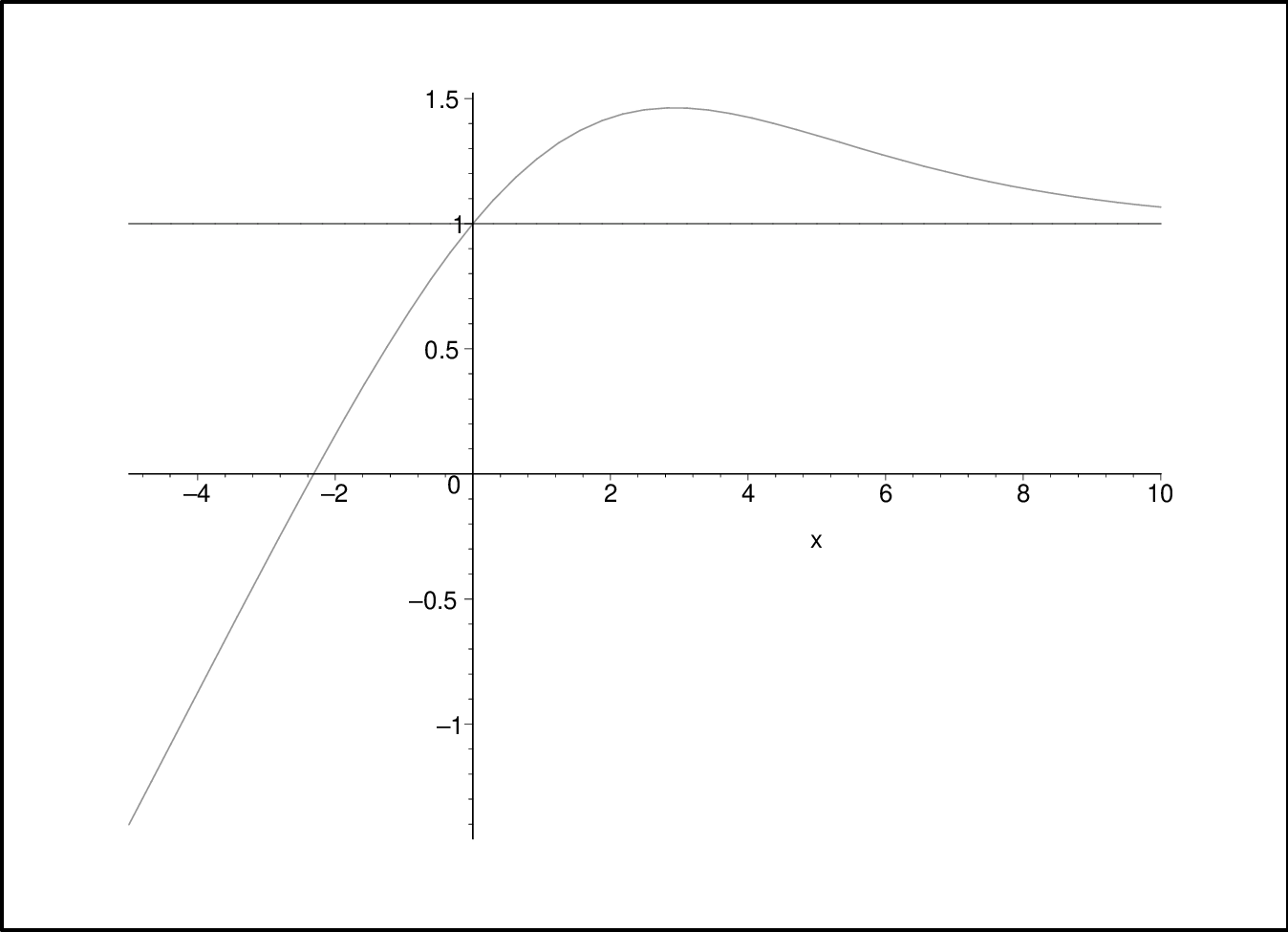}
    \includegraphics[angle=270, scale=0.25]{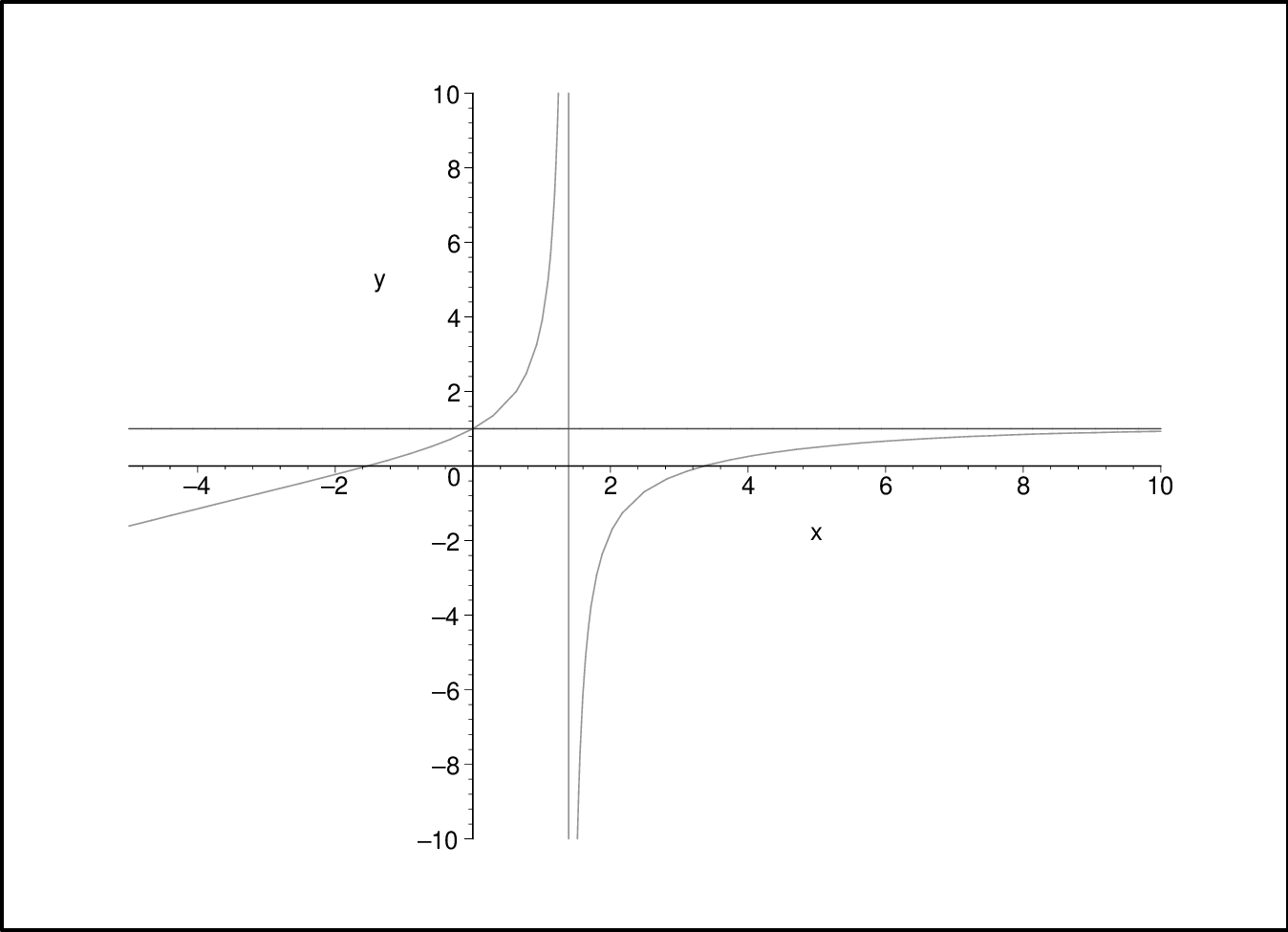}
  \caption{The plot of  solution (\ref{R1}) at $c=1$ (left) and $c=-1$ (right). Horizontal line corresponds to the fixed point $y=1$. Note, that $y(x)$ intersects it at $x=0$.}
\end{center}
\end{figure}
Existence and location of singularity $x_s$ is uniquely determined by the initial conditions $\{x_0,y_0\}$ as follows:
\begin{equation}
x_s=x_0-2\log\left[\frac{2y_0-x_0-2}{2(y_0-1)}\right]. \label{R3}
\end{equation}
If $x_s$, defined via (\ref{R3}) is real, soultion (\ref{R1}) will have a singularity there. In particular, choice of initial conditions lying in $I_1$ and such that $y_0\gg x_0$ will result in singularity (Fig. 2, right) at $x_s\sim x_0$. We would also like to point out that this choice in general leads to the violation of Lipshitz condition that in our case has the form
$$
\left|2(y_2+y_1)-x_0-2\right|\le 2x_0L,
$$
where $L$ is a positive constant.

Note that in order to obtain the aforementioned estimate one can actually avoid the usage of solution  (\ref{R1}), but instead start from  (\ref{Ric}), assuming that  $y\to+\infty$ at $x\to x_s>0$ and integrate the reduced version of (\ref{Ric}) $y'=y^2/x_s$
to get
\begin{equation}
y(x\sim x_s)=\frac{x_s}{x_s-x}. \label{R4}
\end{equation}
(\ref{R4}) can be obtained from (\ref{R1}) with  $C=-2{\rm e}^{-x_s/2}$ by the Taylor expansion around $x=x_s$. Finally, the usage of the initial conditions $\{x_0,y_0\}$ together with (\ref{R4}) results in a crude estimate for $x_s$ (cf. (\ref{R3}), that gets more and more accurate for $y_0\gg 1$:
\begin{equation}
x_s=\frac{y_0x_0}{y_0-1}\to x_0, \label{R*}
\end{equation}
in a good accordance with the results obtained by studying the exact solution (\ref{R1}).

As we have mentioned already, equation (\ref{Abb}) is not the one allowing for explicit integration, hence the study for existence and location of possible singularities has to be performed using the approximate approach that we have tested on the Riccatti equation (\ref{Ric}). The analog
of (\ref{R4}) for (\ref{Abb}) has the form
\begin{equation}
y(x\sim x_s)=\sqrt{\frac{x_s}{2(x_s-x)}}, \label{R5}
\end{equation}
i.e. unlike the discussed singularities of solutions of (\ref{R1}) there might exist at most one left-sided discontinuity point of second order, while for $x>x_s$ the solution is either finite or complex (cf., for example, solution of a simple ODE $y'=y^3$). By analogy with
(\ref{R*}) we'll obtain for (\ref{Abb})
$$
x_s=\frac{2x_0y_0^2}{2y_0^2-1}\to x_0,
$$
when $y_0\gg 1$. This condition together with  $y_0\gg 1$ are also
violating the Lipshitz condition, therefore it is expected for the
singular point to appear near the critical point  $x_s\sim x_0$.
Contrary, the choice $x_0\gg 1$, a $y_0=1+o(1)$ is deemed to
render the nonsingular solution (at least on the right half axes),
by analogy with Fig. 2, left. Note that when $x\to 0$ the
solutions of (\ref{Abb}) has a behaves as $d^ny/dx^n\to 0$ -- a
very interesting property, differing it from solution (\ref{R1})
of Riccatti equation (\ref{Ric}).

For what follows we'll assume that the initial conditions are
chosen in such a fashion that ensures continuity of solution of
(\ref{Abb}) at $x>0$.
\begin{proposition}
Solution of (\ref{Abb}), continuous at  $x\ne 0$ is infinitely differentiable there.
\end{proposition}
{\bf The proof of the proposition.} Continuity of  $y(x)$ is
postulated on the chosen class of initial conditions.
Differentiability of $y(x)$ ar $x\ne 0$ follows directly from
(\ref{Abb}). After $n-1$ differentiation of (\ref{Abb}) w.r.t. $x$
one gets the following:
$$
\frac{d^ny}{dx^n}=\sum_{{\mu=1}}^{k_n} \alpha_{\mu}(x)y^{\mu},
$$
where $k_n=2^n+1$, and coefficients $\alpha_{\mu}(x)$ are continuous and differentiable at $x\ne 0$, which concludes the proof.
\begin{theorem}
Solution of (\ref{Abb}) that belongs to $I_1$  has one and only one maximum on the right half plane $x>0$.
\end{theorem}
{\bf The proof of the theorem.}

First, let us show that existence of extremum of solution $y(x)$ on $x>0$ implies the uniqueness of maximum.

Indeed, assume that $x=x_m>0$ is an extremum of $y(x)$, i.e.
$y'(x_m)=0$. According to (\ref{Abb}) this means that either $y=\pm
1$ or $y_m=y(x_m)=x_m/2$. Since the second derivative at $x_m$ is equal to
$$
y''(x_m)=-\frac{y_m(y_m^2-1)}{x_m^2}.
$$
and $x_m>0$ implies $y_m>0$ it follows that inside of the $I_1$ region $y''(x_m)<0$,
hence $x_m$ can only be the maximum. Uniqueness of maximum then follows straight from the analyticity of $y(x)$ (cf.
Proposition 3).

Now consider the line $y_1(x)=x/2$. According to all the above-said, if
$y(x)$ and $y_1(x)$ intersects at any point $x_m>0$, at this  $y(x)$ suffers a maximum and, moreover, this maximum is unique. Let us now prove that  existence of such intersection point. Indeed, let us assume by contradiction, that  $y(x)$ and $y_1(x)$ does not intersect at $x>0$.
This means either one of two possibilities:
\newline
\newline
A. $y(x)<y_(x)$ at $x>0$ or
\newline
B. $y(x)>y_(x)$ at $x>0$.
\newline

Case A shall be excluded due to the fact that whenever $0<x<2$ we have $y(x)>1$ (region $I_1$), but $y_1(x)<1$. Therefore, we are left with the case B, corresponding to the unbounded increase of function $y(x)$.

It is easy to see that function  $y(x)$ has no oblique asymptote at $x>0$.
Indeed, existence of such asymptote implies the existence of limit of left hand side of (\ref{Abb}), corresponding to the angular coefficient of the asymptote, which contradicts the $x^2$ divergence of the right hand side of  (\ref{Abb}). Thus,  $y(x)$ has to increase fast enough, so that for large values of $x$ the contributions of unities in both parentheses in (\ref{Abb}) will be negligibly small, which, in turn, leads to a simple differential equation $y'=y^3/x$. General solution of this equation contains an unavoidable singularity at a certain point $x_s(C)>0$, whose location depends on the arbitrary integration constant, and for $x>x_s(C)$ the equation has no real solutions. $\Box$

This theorem allows for examinations of the asymptotic behavior of  $y(x)$ at $x\to+\infty$ and at $x\to +0$
provided that $y(x)$ belongs to the region $I_1$.
\begin{proposition}
For $y(x)$ from $I_1$ for $x>0$ the following limits exists:
\begin{equation}
\lim_{_{x\to+\infty}} y(x)=1, \label{lm-1}
\end{equation}
\begin{equation}
\lim_{_{x\to+0}} y(x)=1, \label{lm-2}
\end{equation}
\end{proposition}
{\bf The proof of the proposition.} The
(\ref{lm-1}) is a consequence of the following three facts:
(a) the function $y(x)$ has a maximum at $x>0$; (b) the function
$y(x)$ belongs to $I_1$ and (c) $y=+1$ is a fixed point for equation (\ref{Abb}).

The limit (\ref{lm-2}) is somewhat more interesting. First, note, that despite the fact that $y=1$ is a fixed point, since $x=0$ is a singular point
of (\ref{Abb}) (by analogy with the Riccatti Equation, cf. Fig. 2, left), (\ref{lm-2}) doesn't lead to a contradiction. In order to establish (\ref{lm-2}) it should be kept in mind that, according to the Theorem 2, $y(x)$ is a bounded function, hence for sufficiently small values of  $x$
it is possible to neglect the unity at the right parenthesis of (\ref{Abb}) in order to get the ODE:
$$
y'=\frac{(y^2-1)y}{2x},
$$
whose general solution is
$$
y(x)=\frac{1}{\sqrt{1-c x^2}}.
$$
The (\ref{lm-2}) follows immediately.

All these results can also be justified by the numerical integration; the results of one can be seen on  Fig. 3.

\begin{figure} \begin{center}
  \includegraphics[angle=270, scale=0.35]{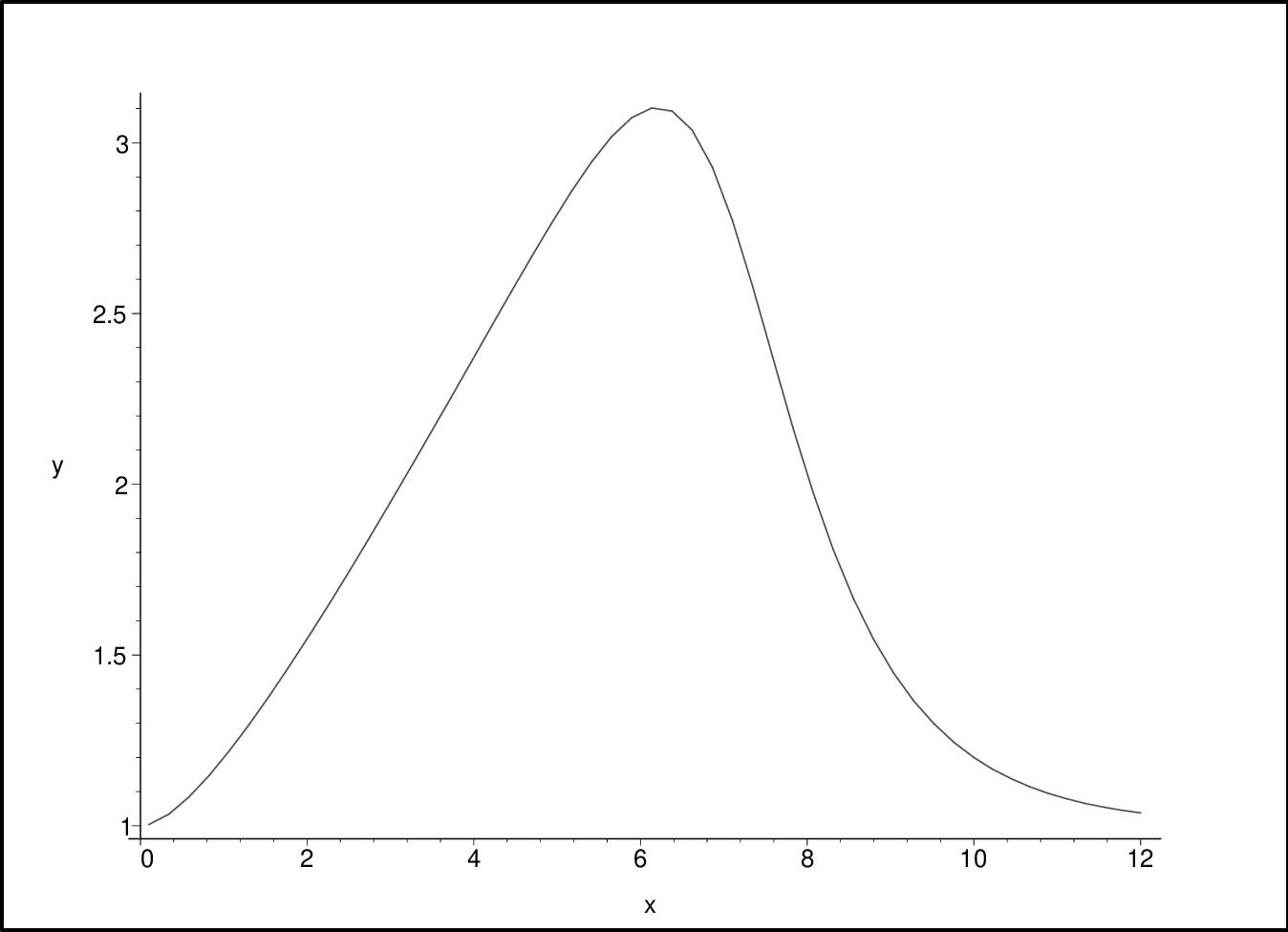}
  \caption{The plot of numerical solution $y(x)$ of the Eq. (\ref{abel}) with initial data
$y(10)=1.2$.}
\end{center}
\end{figure}

The initial value of field $x\gg 1$ so $y\sim 1$ and we have the
''stiff'' equation of state. During the dynamical evolution the
field $x$ decreases. The function $y(x)$ has one point of maximum
$y_{\rm max}=x_m/2$. If $y_{\rm max}>\sqrt{3}$ then this evolution
results in inflation (the region II) (if $y_{\rm max}\gg 1$,
the evolution results instead in slow-rolling regime I). Then the field
decreases and starting out from the $y=\sqrt{3}$ we have the
natural exit from the inflation (the region III) - inflation
indeed ends spontaneously.
\newline
\newline
\noindent\textbf{Remark 9.} The same types of reasonings can be conveyed for another cosmologically and quantum field theory popular model with the potential $V=\lambda\phi^4/4$, where $\lambda>0$. Evidently, all the qualitative properties of its dynamics (inflation's spontaneous start and exit) will remain the same. Moreover, as follows from the analysis, the beginning of the inflation, its exit together with the presence of the slow-rolling regime doesn't depend on the model's physical parameters, be it either scalar particle's mass $m$ (for quadratic potential) or the
coupling $\lambda$ for the $\phi^4$ model. This demonstrate the universality of inflation, at least for the class of the power potentials.
\newline
\newline
\noindent\textbf{Remark 10.} We have used above the analyticity of solutions of  (\ref{Abb}) at $x>0$. This condition will be of crucial importance in the studies on the (becoming very popular recently) effect of the crossing of the phantom divide line. In a series of works (cf., for example, \cite{int-4}, \cite{AKam}) it has been shown that during the cosmic evolution it is possible to for kinetic term to become negative valued. In order for this to happen there should exist a region where kinetic term becomes equal to zero, i.e. ${\dot\phi}=0$. This implies the existence of a point $x_*$, such that $W(x_*)=V(x_*)$. Comparing this with  (\ref{theW}) it is possible to show that  $x_*$ has to be a second order break point for the function $y(x)$. Hence, for the discussed power potentials the effect of the crossing of the phantom divide line is out of the question.

\subsection*{6. The B\"{a}cklund  auto-transformations for the Abel equation}

Throughout this paper we have used the Abel equation for construction of the
general solution of the  Einstein-Friedmann equations for the models of
universe filled with scalar field with the given potentials. In this
Section we'll show that one can use the established connection to study some
interesting  mathematical properties of  the Abel equation itself.

\begin{theorem}
The Abel equation (\ref{abel}) admits a set of B\"{a}cklund
auto-transformations:
\begin{equation}
\chi(x)\to\chi^{(1)}(x),\qquad y(x)\to y^{(1)}(x), \label{BB}
\end{equation}
satisfying the following properties:
\newline
\newline
{\bf Property 1.} If $y(x,C)$ is a general solution of the
(\ref{abel}) with given $\chi(x)$, then  (\ref{BB}) applied to $y$
and $\chi$ generates a general solution
$$
\chi(x)\to\chi^{(1)}(x),\qquad y(x,C)\to y^{(1)}(x,C),
$$
of a transformed equation
\begin{equation}
\left(y^{(1)}\right)'=-\frac{1}{2}\left(\left(y^{(1)}\right)^2-1\right)\left(1-\left(\chi^{(1)}\right)'y^{(1)}\right),
\label{AA-til}
\end{equation}
{\bf Property 2.} This transformations can be applied arbitrarily many times; after the  $n$-th step the n-times transformed function $y^{(n)}\left(x,C\right)$ will contain an arbitrary constant $C$, while $\chi^{(n)}=\chi^{(n)}(x)$.
\newline
\newline
{\bf Property 3.} The transformations can be inverted by setting the integration constant equal to zero. For example, for $n=2$
$$
\chi(x)\to\chi^{(2)}=\chi^{(2)}\left(x,C^{(1)}=0\right)=\chi(x),
$$
where $C^{(1)}$ is a certain intermediate parameter.
\end{theorem}
{\bf The proof of the Theorem 3.} It was shown that using exact
solution of the Abel equation (\ref{abel}) $y(x,C)$ it is possible
to find the exact solutions $x(t,t_0,C)$ (i.e. $\phi(t,t_0,C)$)
and $a(t,t_0,C)$ of the Einstein-Friedmann equations in the flat
space-time. Define function $\psi(t,t_0,C)=a^3(t,t_0,C)$. One
can show that $\psi$ is the solution of the Schr\"{o}dinger
equation \cite{chervon}, \cite{int-4} (with the notation
$\ddot\psi=d^2\psi/dt^2$):
\begin{equation}
{\ddot\psi}=9U_{_C}(t)\psi, \label{schrod}
\end{equation}
where $U_{_C}(t)=V(x(t,C))$.
\newline
\newline
\noindent\textbf{Remark 11.} $U_{_C}(t)$, unlike to  $V(x)$ does depend on constant $C$ (for simplicity we'll omit constant  $t_0$ for everything that follows). In fact, using equation (\ref{newphi}) it is possible to present $t$ as a function of $x$ and $C$:
\begin{equation}
t=t(x,C)\mp\frac{1}{6}\int dx\,\frac{\sqrt{W(x,C)}}{W'(x,C)}.
\label{txc}
\end{equation}
Inverting the (\ref{txc}) (this can be done explicitly, as we will show below) we'll get $x=x(t,C)$, i.e.
$V(x)=U_{_C}(t)$. This observation gives rise to what we'll call the {\bf reparametrization invariance} of function  $U$:
\begin{equation}
U_{_C}(t)=U_{_C}(t(x,C))=V(x). \label{repar}
\end{equation}
This property will prove to be of utmost importance for the proof of the Property 1.

Equation (\ref{schrod}) admits Darboux transformation
\cite{D1882}, \cite{Salle}:
\begin{equation}
U_{_C}(t)\to
U^{(1)}_{_C}(t)=U_{_C}(t)-\frac{2}{9}\frac{d^2}{dt^2}\log\psi(t,C),
\label{DT-1}
\end{equation}
\begin{equation}
\psi(t,C)\to\psi^{(1)}(t,C,C^{(1)})=\frac{1}{\psi(t,C)}\left(1+C^{(1)}\int
dt\,\psi^2(t,C)\right), \label{DT-2}
\end{equation}
where $C^{(1)}$ is a new arbitrary constant of integration.
Another one of new integration constants in (\ref{DT-2}) is
introduced multiplicatively and cancels for all the formulas
below. Using the equations (\ref{fried}) it is possible to define
a new field variable
\begin{equation}
x^{(1)}=x_0^{(1)}\pm 2\int
dt\,\sqrt{\frac{d^2}{dt^2}\log\frac{1}{\psi^{(1)}}}, \label{Oh2}
\end{equation}
as a function of time  $x^{(1)}=x^{(1)}(t,C,C^{(1)})$. Inverting this relation, we obtain $t=t(x^{(1)},C^{(1)})$. Of course, the inversion of function in general is a highly nontrivial task, but the one easily accomplished in our particular case.
\begin{proposition}
The relationship $x=x(t)$ can be explicitly inverted (as t=t(x)) for all explicitly known solutions of equation
(\ref{abel}) (or (\ref{fried})).
\end{proposition}
In order to justify this statement one has to refer to (\ref{txc}), which is a result of inversion of
$x=x(t)$, defined by (\ref{Oh2}) with the substitution
$\psi^{(1)}\to\psi$, $x_0^{(1)}\to x_0$.

Using the relation $t=t(x^{(1)},C,C^{(1)})$ it is possible to calculate the quantities $U^{(1)}_{_C}(t)=U^{(1)}_{_C}(t(x^{(1)},C,C^{(1)}))$.
Assuming $C^{(1)}=C$, one gets
$U^{(1)}_{_C}(t)=U^{(1)}_{_C}(t(x^{(1)},C))$. Due to the reparametrization invariance  (\ref{repar}):
\begin{equation}
V^{(1)}(x^{(1)})=U^{(1)}_{_C}(t(x^{(1)},C)). \label{repar-1}
\end{equation}
I.e., the  {\em reparametrization invariance} assures that a new potential $V^{(1)}$ depends on the new independent variable $x^{(1)}$, but not on  $C$. This is by all means true also for  $\chi^{(1)}=\chi^{(1)}(x^{(1)})$.

On the next one is to calculate the quantity
$$
W^{(1)}(x^{(1)},C)=\left(H^{(1)}(t(x^{(1)},C))\right)^2,
$$
where
$$
H^{(1)}(t)\equiv \frac{1}{3}\frac{d}{dt}\log\psi^{(1)},
$$
and $\psi^{(1)}$ shall be defined from (\ref{DT-2}) with regards of the substitution $C^{(1)}=C$. Finally, using the formula
(\ref{theW})
$$
W^{(1)}(x^{(1)},C)=V^{(1)}(x^{(1)})\theta^2(y^{(1)}(x^{(1)},C)),
$$
where $\theta^2(y)$ is defined by (\ref{thetta}) one derives the thought-after ''dressed'' solution of Abel equation
\begin{equation}
y^{(1)}(x^{(1)},C)=\sqrt{\frac{W^{(1)}(x^{(1)},C)}{W^{(1)}(x^{(1)},C)-V^{(1)}(x^{(1)})}}.
\label{akrobatypidory}
\end{equation}
Formulas (\ref{repar-1}) and (\ref{akrobatypidory}) defines
the B\"{a}cklund  auto-transformations for the Abel equation
(\ref{abel}) and proves the Property 1. The proof of Property 2 is based on the usage of n-fold consequent Darboux transformations and the Crum  determinant formulas \cite{Crum}, \cite{BLP}, \cite{TemaLera}. The Property 3 can be checked directly: it is enough to let $C^{(1)}=0$ (but before one lets this parameter be equal to $C$, because at this step $C^{(1)}$ and
$C$ are still being treated as the independent constants) in (\ref{DT-2}) and perform the second Darboux transformation:
$$
U^{(1)}_{_{C}}\to
U^{(1)}_{_{C}}-\frac{2}{9}\frac{d^2}{dt^2}\log\psi^{(1)}(t,C,C^{(1)}=0)=U_{_{C}}(t),
$$
Q.E.D. $\Box$

 \vspace{.2in} \noindent{\large {\bf Acknowledgements:}}

 Research has been partially supported by
EPSRC Grant GR/S13682/01 and  by the Russian Foundation for Basic
Research (Grants No. 08-02-91307-${\rm IND}_{\rm a}$, No.
09-05-00446ba).

\medskip
\noindent {\bf Authors:}\\ \hfill \\
{\tt Artyom V. Yurov:} Department of Theoretical Physics, Russian
State University of Immanuel Kant, Aleksandra Nevskogo 14,
Kaliningrad 236041, Russia; e-mail {\em artyom\_yurov@mail.ru}

\medskip \noindent
{\tt Valerian A. Yurov:} Department of Mathematics, University of
Missouri, Columbia 65201, US; e-mail {\em
valerian@math.mussouri.edu}

\end{document}